\documentclass[letterpaper,twocolumn,10pt]{article}

\newif\ifieeestyle
\ieeestylefalse

\ifieeestyle
    \usepackage[utf8]{inputenc}
    \usepackage{xcolor}
\fi

\usepackage{amsmath,amssymb,amsfonts}
\usepackage[
    ruled,
    noend,
    noline
]{algorithm2e}
\usepackage[notransparent]{svg}
\svgpath{{./fig/img/}}
\usepackage{graphicx}
\graphicspath{{./fig/img/}}
\usepackage{textcomp}
\usepackage{enumitem}
\usepackage{booktabs}
\usepackage{multirow}

\ifieeestyle
    \usepackage[hyphens]{url}
    \usepackage[hidelinks]{hyperref}
    \hypersetup{breaklinks=true}
    \usepackage{cite}

\fi

\makeatletter
\newcommand{\escapeus}{\begingroup\@makeother\_\@escapeus}
\newcommand*{\@escapeus}[1]{#1\endgroup}
\makeatother

\def\BibTeX{{\rm B\kern-.05em{\sc i\kern-.025em b}\kern-.08em
    T\kern-.1667em\lower.7ex\hbox{E}\kern-.125emX}}

\newif\ifanon
\anontrue

\ifanon
\fi

\usepackage{style/usenix-2020-09}
\usepackage{xcolor}
\usepackage{listings}
\usepackage{nopageno}

\begin{document}

\date{}

\title{\Large \bf Fuzzing BusyBox: Leveraging LLM and Crash Reuse for Embedded Bug Unearthing}

\author{
{\rm Asmita$^1$, Yaroslav Oliinyk$^2$, Michael Scott$^2$, Ryan Tsang$^1$,} \and
{\rm Chongzhou Fang$^1$, Houman Homayoun$^1$} \\ \\
$^1$University of California, Davis
\quad $^2$NetRise
}



\maketitle

\begin{abstract}

  BusyBox, an open-source software bundling over 300 essential Linux commands into a single executable, is ubiquitous in Linux-based embedded devices. Vulnerabilities in BusyBox can have far-reaching consequences, affecting a wide array of devices. This research, driven by the extensive use of BusyBox, delved into its analysis. The study revealed the prevalence of older BusyBox versions in real-world embedded products, prompting us to conduct fuzz testing on BusyBox. Fuzzing, a pivotal software testing method, aims to induce crashes that are subsequently scrutinized to uncover vulnerabilities. Within this study, we introduce two techniques to fortify software testing. The first technique enhances fuzzing by leveraging Large Language Models (LLM) to generate target-specific initial seeds. Our study showed a substantial increase in crashes when using LLM-generated initial seeds, highlighting the potential of LLM to efficiently tackle the typically labor-intensive task of generating target-specific initial seeds. The second technique involves repurposing previously acquired crash data from similar fuzzed targets before initiating fuzzing on a new target. This approach streamlines the time-consuming fuzz testing process by providing crash data directly to the new target before commencing fuzzing. We successfully identified crashes in the latest BusyBox target without conducting traditional fuzzing, emphasizing the effectiveness of LLM and crash reuse techniques in enhancing software testing and improving vulnerability detection in embedded systems. Additionally, manual triaging was performed to identify the nature of crashes in the latest BusyBox.

\end{abstract}

\section{Introduction}
\label{sec:intro}

The proliferation of IoT (Internet of Things) devices continues unabated, with a reported 16\% growth rate propelling the global count to 16.7 billion, according to IoT Analytics \cite{iot-analytics}. This remarkable expansion in the IoT ecosystem raises significant cybersecurity concerns. Embedded devices occupy a central role in IoT security, as vulnerabilities within them can jeopardize an entire system's security. Because of the role it plays in governing most aspects of a system's behavior, firmware is of particular importance. Often, firmware is comprised of numerous third-party software components that can be reused across various products, thereby amplifying concerns regarding their vulnerability; a single flaw could potentially affect multiple disparate devices that rely on shared components. Consequently, continuous analysis of these components is imperative. 

\subsection{Motivation}
\label{subsec:motivation}

Firmware can be broadly classified into three categories: those based on modified generic operating systems (OS) like Linux, those based on real-time (RTOS) or custom operating systems, and those that do not have a formal operating system (non-OS or bare-metal).
Each of these categories poses distinct challenges when it comes to security assessment \cite{WYCINWYC} and often require different approaches. To that end, we focus our attention in this work on the largest subclass of OS-based firmware: Embedded Linux.

Embedded Linux-based firmware often utilizes various application-level software components, including BusyBox, Lighthttpd, Dropbear, SQLite, OpenSSL, telnet server, and various file system utilities. Consequently, there are multiple potential attack surfaces, warranting continued exploration and research. As a critical set of commonly used utility programs in embedded Linux, BusyBox \cite{busybox} is a component of particular interest. It provides over 300 common Unix utilities within a single lightweight and compact executable, making it indispensable for resource-constrained Linux-based embedded devices. There are many IoT and OT(Operational Technology) devices running BusyBox, including remote terminal units (RTUs), human-machine interfaces (HMIs), and many others that are running on Linux. However, despite its many advantages, it can also present considerable risk, as it is often used with elevated privileges and provides multiple utilities that handle user input, which attackers have been able to exploit. 14 vulnerabilities were found in Busybox in 2021, some of which had the potential of remote code execution or denial of service attacks \cite{Constantin2021-BusyBoxFlawsHighlight}. Despite this, in our investigation we have identified several real-world products that continue to use older versions of BusyBox that contain known vulnerabilities. 

Our research questions for this work can be summarized as follows:

\begin{enumerate}[label=\textbf{Q\arabic*:},noitemsep]
    \item How widespread are variants of BusyBox and how can similar vulnerabilities across variants be efficiently identified?
    \item How can we leverage LLMs to improve fuzz testing on embedded Linux utility programs like those in BusyBox?
\end{enumerate}

\subsection{Contributions}
\label{subsec:contribution}


Fuzzing is a well-recognized software testing technique for uncovering vulnerabilities, but its effectiveness varies depending on the chosen target, each of which can present unique challenges. In this work, we propose and implement two techniques to assist software testing for embedded linux. First, we leverage \textbf{LLM-based seed generation}, in which we utilize commercial large language models (LLMs) to generate the initial input seeds for mutation-based, coverage-guided fuzzing. In doing so, we take advantage of LLMs' inherent capability to generate high-quality structured inputs that adhere to the input grammar of a target. Second, we employ a \textbf{crash reuse} strategy to identify crashes across variants of a software component present in different targets. This strategy is based on the intuition that an input that triggers a crashing vulnerability on one variant of a program is likely to trigger a crash on different variant. This allows us to more efficiently determine if the same vulnerability is present on multiple program variants without performing fuzzing, thus saving significant time. When we mention a variant of a software component, we are referring to identical software components with varying version numbers or architectures or compiler optimization, or any custom modification by developers.


As a proof-of-concept, we demonstrate these techniques with AFL++ on BusyBox. This research was done in collaboration with NetRise's \cite{NetRise} firmware security division. We sourced BusyBox ELFs from real-world embedded products collected from the company's proprietary firmware dataset, which had been constructed using in-house extraction tools. These ELF binaries were fuzz tested without altering the target's source code or compilation process, using AFL++ in QEMU mode. 

We evaluated the first technique, LLM-based seed generation, by comparing control runs that used randomly generated initial seeds, to experimental runs that used initial seeds generated using OpenAI's GPT-4 LLM API \cite{OpenAI}. We observed a significant increase in crashes obtained when using LLM-generated seeds, demonstrating the potential for improving vulnerability detection.


We then used the crash results accumulated from these experiments to evaluate the second technique, crash reuse, by testing them against the latest version of BusyBox (v1.36.1 at the time of writing). Since the accumulated crashes corresponded to older versions of BusyBox, any reused inputs that still cause crashes are likely due to the continued presence of the same vulnerability, which allow us to discover crashes in the latest version without fuzzing it explicitly. We also fuzz tested the latest version and conducted a comparative analysis. The results of our validation highlighted the effectiveness of crash reuse with respect to time and resource efficiency.


In the context of open-source software components, collecting crashes can be enhanced by instrumenting the source code and improving crash identification. These collected crashes can then be applied to the variant of that software component present in different targets, even in cases involving black-box testing. This is because, in many instances, the variant of software component may reappear in different target. Therefore, reusing existing crashes from a particular software component for testing its variant on different targets can be advantageous. This approach streamlines the provision of crash data directly to the target before the commencement of fuzzing, resulting in substantial time savings. 
\begin{figure}[t]
	\centering
	\includegraphics[width=\columnwidth]{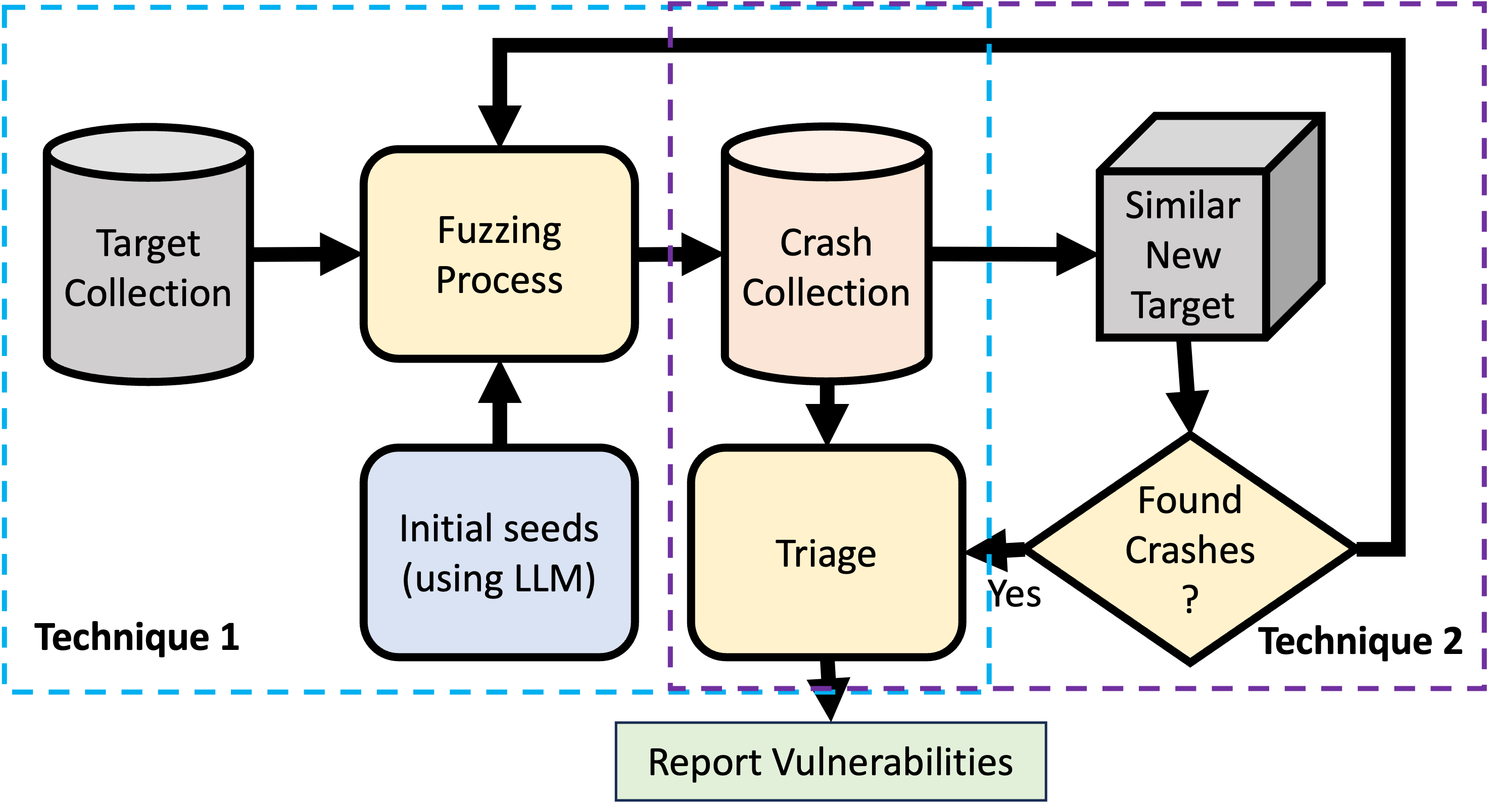}
	\caption{Proposed work pipeline}
	\label{fig:pipeline}
\end{figure}

Finally, we conducted crash analysis for the latest BusyBox version(v1.36.1), compiled from the BusyBox source code. Figure \ref{fig:pipeline} shows the overall pipeline of the proposed techniques. Our contributions can be summarized as follows:

\begin{itemize} 
    \item We identify versions of BusyBox still in use in commercial embedded devices, emphasizing the need to update them.
    \item We implement \textbf{LLM-based seed generation} to enhance fuzzing by utilizing LLM for target-specific initial seed generation, leading to faster and more efficient seed generation, more crashes, and more options for triaging to identify vulnerabilities. We developed an automation script to perform fuzzing on a large batch of BusyBox targets without manual intervention, which will be made open source.
    \item We propose \textbf{crash reuse} as a first-pass bug-finding strategy, in which we reuse crashing inputs for different program variants to quickly find duplicate vulnerabilities, even under black-box testing conditions.
    \item We identify crashes in the \texttt{awk} applet of latest BusyBox version and conduct manual crash triaging to determine whether they originated from BusyBox or dependencies in the underlying libraries. We later applied these techniques to other applets including \texttt{dc}, \texttt{man}, and \texttt{ash}.
\end{itemize}

\section{Background}
\label{sec:background}
In this section, we provide a brief overview of some of the tools and techniques relevant to this paper.

\subsection{BusyBox}
BusyBox \cite{busybox} is a single binary executable for several Unix-based utilities designed mainly for resource-constrained embedded devices. It is open source, lightweight, compact, and has a small footprint. It allows manufacturers to include essential Linux utilities without significantly increasing the firmware size. Moreover, it is highly customizable. It can be configured to include only the specific utilities required for the embedded system's functionality. 

However, it has associated potential security risks. BusyBox often runs with elevated privileges as various tasks require root access, hence a potential risk of privilege escalation. It also provides a variety of commands that accept user input. If these commands are not properly sanitized, it can lead to command injection \cite{command-injection}, buffer overflow \cite{buffer-overflow}, and other vulnerabilities \cite{busybox-vulnerabilties}. For instance, \texttt{awk} applet is used for text processing and data manipulation tasks. It can be used to do privileged read or write outside a restricted file system as it writes and reads data to and from files. Similarly, other applets may possess security risks when not configured or used correctly. Moreover, as BusyBox aims to be small and lightweight, it may lack some of the security features, or the issues might also occur because of its other external dependencies. Given its critical role in many embedded systems and its potential security risks, securing BusyBox through proper configuration, regular updates, code reviews, and security assessments are essential to maintaining the overall security and reliability of embedded Linux-based firmware.

\subsection{Fuzzing}
Fuzzing is one of the widely used methods for software testing. It leverages coverage feedback and aims to identify if any crash occurs, which is then analyzed to identify the vulnerabilities. Fuzzing can be described as input testing, which is semi-randomized due to the impossibility of exhaustive input testing in most cases. In vanilla fuzzing, a program input such as a variable, buffer, stream, or file is selected as the target of the fuzzing process. The fuzzing engine, also known as the fuzzer, generates input based on a set of rules or strategies, typically using some element of randomness, and feeds it to the program. The program is then executed with this input and observed for coverage information. The process is repeated with a newly generated input. This process can be repeated for as long as the analyst desires. During the execution phase, the program may crash or trigger some other error. If this occurs, the input that caused the issue is saved for later root cause analysis. This type of fuzzing is known as black-box fuzzing because the fuzzer is unaware of the program's internals.
In contrast to black-box fuzzing, white-box fuzzing assumes complete knowledge of the program's internal structure. It allows input generation strategies to use techniques such as taint analysis or symbolic execution, which require understanding the program's syntax. In between white and black-box fuzzing is grey-box fuzzing, where there is only partial access to information about the program's internal state. In this approach, code coverage, which refers to the number of lines of code or basic blocks that are actually executed, is often used to direct input generation. Program code must first undergo instrumentation to gather coverage information, in which additional code is inserted into the program to act as a marker and collect runtime data. 

\subsubsection{AFL/AFL++}
American fuzzy lop (AFL) is a coverage-guided, grey-box fuzzer that employs compile-time instrumentation and several different algorithms to efficiently fuzz programs \cite{AFL}. At a high level, AFL tracks branch coverage between basic blocks (edges) for a generated input file, keeping track of all the edges that the input file yields. Whenever a new input file is created, AFL checks if it leads to previously unseen edges, saves the file, and uses it to inform future mutations. AFL uses deterministic (such as sequential bit flips and sequential addition of small and interesting integers) and nondeterministic (such as a stacked sequence of randomized operations with equal probability) mutation strategies \cite{AFL++}. 
AFL\cite{AFL} has been extensively used in industry and academic research. However, the original developer ceased active development on AFL in 2017, leading to the development of a community-driven successor, AFL++, in 2019. AFL++ features new enhancements, fuzzing strategies, and performance improvements \cite{AFL++}. 

AFL++ provides detailed documentation for all its supported features. If the source code is available, the first stage involves modifying and compiling the target using the AFL++ compiler. Otherwise, when the source code is unavailable, AFL++ QEMU \cite{QEMU} performs internal instrumentation at runtime. The next stage is to provide the initial seed corpus for fuzzing the target, followed by the actual fuzzing stage and crash triaging. 


\subsection{LLM in Fuzzing}
Large language models (LLMs) are known for their prowess in natural language understanding and text generation. Its primary focus has been developing AI models and technologies, such as the GPT (Generative Pre-trained Transformer) series \cite{GPT}, for various applications, including natural language understanding, generation, and automation. While LLM's work may not be directly related to fuzz testing, there are potential intersections between AI and security, where AI-powered tools and techniques could be used to enhance security testing practices \cite{AI_cyber}, including fuzzing. AI can assist in automated test case generation, identifying patterns in code more likely to contain vulnerabilities, and analyzing code patterns, among other tasks. Some existing work leverages LLM for fuzzing. In a recent blog by Google open source security team \cite{OSS-Fuzz-LLM}, an LLM-aided fuzzing is proposed. OSS-Fuzz \cite{OSS-fuzz} is integrated with the LLM to assess its potential to generate new fuzz targets effectively. OSS-Fuzz's Fuzz Introspector tool identifies and sends the under-fuzzed and high-potential part of the project code to the evaluation framework. A prompt is created by the evaluation framework, which embeds project-specific information. LLM subsequently uses the prompt to write a new fuzz target. The newly generated fuzz target is shared back with the evaluation framework, which executes it and monitors for any change in the code coverage. In case of any compilation failure, it prompts the LLM to revise the fuzz target addressing the compilation failure. This comprehensive approach shows how LLMs can be harnessed to completely automate the fuzz testing process of OSS-Fuzz, contributing to its efficiency and efficacy. Other related works include ChatFuzz \cite{Hu2023-AugmentingGreyboxFuzzing}, an LLM-based fuzzer that enhances the quality of format-conforming inputs for fuzzing, and Fuzz4All \cite{Xia2024-Fuzz4AllUniversalFuzzing}. This system utilizes LLM for input generation and mutation, producing diverse and realistic inputs for various programming languages. Fuzz4All mainly targets systems that accept programming languages as input. Additionally, FuzzGPT \cite{FuzzGPT} uses LLM as a fuzzer to test deep-learning libraries. Similarly, as discussed in Section \ref{sub-teq1}, our work leverages LLM to generate initial seeds for fuzzing BusyBox. 

However, what sets our approach apart is that we exclusively employ LLM for the initial seed generation stage targeting embedded application, unlike other works where LLM is integrated into the entire fuzzing or mutation process. As previously discussed, fuzzers rely on initial input seeds for the target. While random seeds can be used, the performance significantly improves when the initial seeds align with the expected inputs for the target. Common initial test cases for various inputs, such as images, videos, PDFs, XML, and HTML, are readily available. However, in some cases, generating these initial seeds can be challenging. In such situations, leveraging LLM allows us to generate initial seeds simply by providing information about the target type.



\section{Related Works}
\label{sec:related}

\subsection{Command Line Fuzzing}
\label{subsec:command-line-fuzzing}
CLI programs were the first to be subjected to what is now the fuzz testing technique with Miller et al.’s studies on the reliability of UNIX utility programs \cite{Miller1990-EmpiricalStudyReliability, Miller1998-FuzzRevisitedReExamination, Miller2007-EmpiricalStudyRobustness}. Miller et al. recently repeated the classic fuzz test for  a number of Unix utilities on Linux, FreeBSD, and MacOS \cite{Miller2022-RelevanceClassicFuzz}, demonstrating the relevance of classical fuzz testing for command line utilities even now. They used random input generation technique. 

Since the original studies, fuzz testing has grown into a flourishing area of research, with CLI utilities continuing to be a major target \cite{AFL, Bohme2019-CoverageBasedGreyboxFuzzing, Pham2021-SmartGreyboxFuzzing, OSS-fuzz, Zhu2021-RegressionGreyboxFuzzing, Zhu2022-CSIFuzzFullSpeedEdge}. Significant improvements have been made by considering grammars to define input formatting constraints during seed generation \cite{Wang2017-SkyfireDataDrivenSeed, Godefroid2017-LearnFuzzMachine, Aschermann2019-NAUTILUSFishingDeep, Wang2019-SuperionGrammarAwareGreybox, Blazytko2019-GRIMOIRESynthesizingStructure}, which is also relevant to fuzzing command line arguments in particular. 

Song et al. \cite{Song2020-CrFuzzFuzzingMultipurpose} noted that most off-the-shelf fuzzers do not deal well with conditional option parameters and introduced CrFuzz to more efficiently explore multi-purpose programs by adding input validity prediction to existing fuzzers. Gupta et al. \cite{Gupta2022-CLIFuzzerMiningGrammars} took a different approach to enable systematic testing of command line options by defining a grammar for valid sequences of options and arguments based on the \texttt{getopt} function. Zhang et al. \cite{Zhang2023-FuzzingConfigurationsProgram} propose ConfigFuzz, which transforms the program under test to treat configuration parameters as potential fuzzing inputs. Wang et al. \cite{Wang2023-CarpetFuzzAutomaticProgram} propose CarpetFuzz to extract command line option relationships from documentation using NLP to improve the efficiency of fuzzing different option combinations.

The introduction of LLMs is already making waves in the fuzzing community. OSS-fuzz \cite{Huang2024-LargeLanguageModels, OSS-fuzz} has begun experiments to explore fuzzing new targets with LLMs \cite{OSS-Fuzz-LLM}. Huang et al. \cite{Huang2024-LargeLanguageModels} survey a number of LLM-based fuzzers, at least 4 of which appear to be directly applicable for CLI tools \cite{Hu2023-AugmentingGreyboxFuzzing, Yang2023-WhiteboxCompilerFuzzing, Meng2024-LargeLanguageModel, Xia2024-Fuzz4AllUniversalFuzzing}. Other works have employed LLMs for the purposes of direct input generation \cite{Yan2023-ParaFuzzInterpretabilityDrivenTechnique, Zhang2023-UnderstandingLargeLanguage}, mutation \cite{Deng2023-LargeLanguageModels}, and seed generation \cite{Meng2024-LargeLanguageModel, Hu2023-AugmentingGreyboxFuzzing}. LLMs have also been employed to generate inputs that have irregular or unique input grammars and semantics \cite{Qiu2023-CHEMFUZZLargeLanguage, Liu2023-TestingLimitsUnusual}. These, however, were applied to software meant for general-purpose systems, and none of these have explored the application to fuzzing in embedded environments like BusyBox.

\subsection{Fuzzing in Embedded Environments}
\label{subsec:fuzzing-embedded-envs}

To our knowledge and upon conducting thorough research, we have found no dedicated paper addressing the topic of BusyBox fuzzing. However, there have been various blogs and articles discussing this subject. For instance, Claroty and JFrog \cite{jfrog-busybox-vuln} identified 14 vulnerabilities in BusyBox version 1.34.0. It's worth noting that other related works often focus on uncovering vulnerabilities in specific targets. However, fuzz testing in Linux-based embedded systems is an active area of research.

In an early application of fuzzing in embedded systems, Sim et al. \cite{Sim2011-FuzzingOutofmemoryKiller} applied black-box fuzzing to the Out-Of-Memory Killer process on embedded Linux, which revealed a number of failure modes that would cause the kernel to remain in the Out-Of-Memory state and unresponsive. They implemented an adaptive random approach to input generation that reduced the number of inputs necessary to expose failures.

Du et al. \cite{AflIoT} presents AFLIoT, an on-device fuzzing framework for Linux-based IoT firmware. It involves binary-level instrumentation techniques. It leverages an AFL fuzzer. It has two phases, namely, the instrumentation and fuzzing phase. The implementation involves storing the fuzzer and the instrumented program on the device. AFL fetches the execution information of the target binary by shared memory. It sets up a shared memory before forking the target program and then maps it into its memory space. The instrumented code is responsible for keeping track of basic block transition, which AFL then analyzes to assess the value of the test case. The authors have also implemented the input redirection mechanism between the fuzzer and the target network daemon program. AFLIoT identified 437 unique crashes, out of which 95 were newly found. It was tested on 13 binary programs. The authors have evaluated both benchmarks and real-world IoT devices.  

Zheng et al. \cite{greybox-fuzzing-iot-devices} proposed EQUAFL, an efficient greybox fuzzing for Linux-based IoT devices using enhanced user-mode emulation. It automatically sets up the execution environment to execute embedded applications. It first executes the application under full-system emulation and observes the points where the target may crash or stuck during user-mode emulation. Then, depending on the observed information, it migrates the needed environment for user-mode emulation. It supports the replay of system calls of network and resource management behavior. The approach involves using lightweight program instrumentation to collect execution feedback of the program under test (PUT), such as code coverage, to guide the entire testing process. The authors propose to observe the dynamic configuration file generation and NVRAM configurations with process awareness, network behaviors with state awareness, and other information such as launch variables and process resource limits using several heuristics. The authors conducted experiments on several real-world IoT devices and demonstrated that their approach outperforms existing techniques regarding code coverage, vulnerability discovery, and execution time. 

All the work described above has involved a significant effort to develop improved techniques for enhancing the security assessment of respective targets. Other works that involve fuzzing embedded Linux have focused on enabling effective fuzzing through emulation and or rehosting \cite{Srivastava2019-FirmFuzzAutomatedIoT, Liu2021-FirmGuideBoostingCapability, Jiang2021-ECMOPeripheralTransplantation}, which is an open problem on its own. Our work is intended to continue improving fuzzing techniques and strategies for embedded Linux targets by utilizing LLMs and crash reuse.

\section{Experiment}
\label{sec:implement}

In this section we outline the sample collection process, our implementation of LLM-based seed generation, using the Awk applet as an example, and our experimental procedures for crash reuse and analysis.

\begin{figure*}[t]
	\centering
	\includegraphics[width=\textwidth]{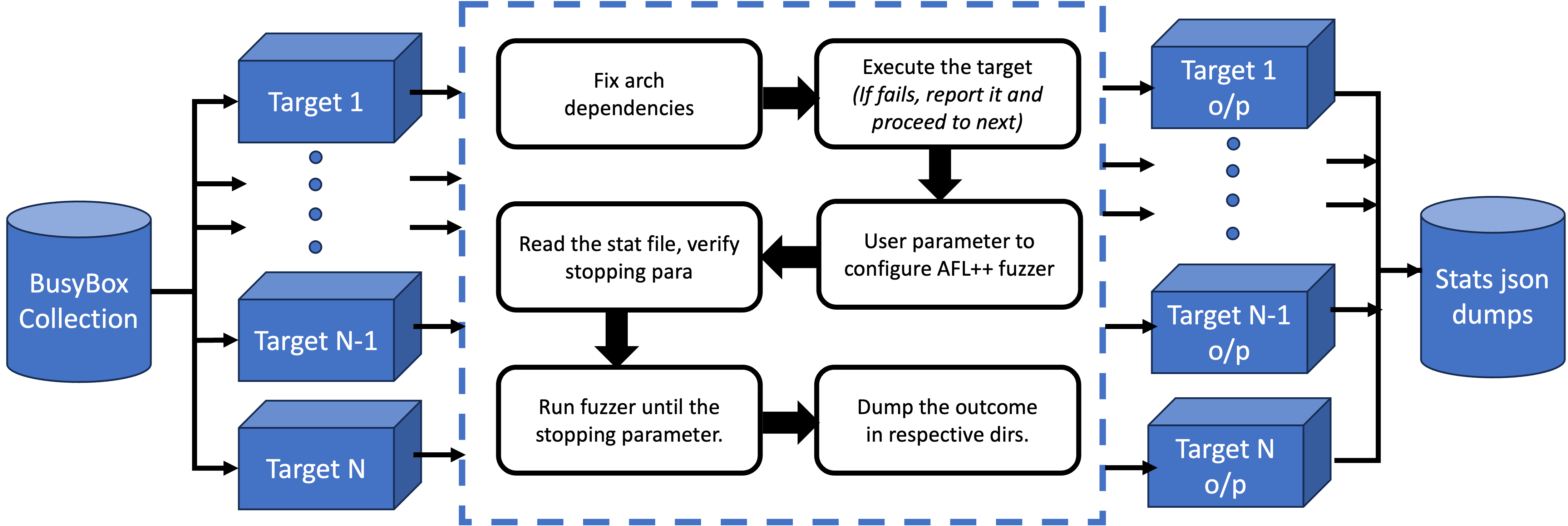}
	\caption{Automation Framework Workflow \textit{(N.B: Target o/p contains crashes, queues along with other stats related to fuzzing)}}
	\label{fig:automation-framework}
\end{figure*}

\subsection{Analyzing BusyBox Versions in Real-World Products}
\label{sub-version}

As mentioned in Section \ref{subsec:motivation}, BusyBox is widely used in Linux-based embedded devices. In order to understand the scope of impact that vulnerabilities in BusyBox variants might have (RQ1), we conducted a brief investigation on the prevalence of older versions of BusyBox within real-world products. To achieve this, we harnessed a proprietary firmware dataset provided by the company. This dataset was curated using the company's platform, which was employed to extract the collected firmware samples. Within these extracted filesystems, we identified BusyBox ELF binaries. We identified \texttt{293} BusyBox ELF binaries distributed across various real-world firmware binaries within the small realm of the provided dataset.

Within the scope of our analysis, we focused on approximately 80 ELFs from ARM\_32 and x86\_64 architectures with 30 BusyBox variants across them. Each of these BusyBox ELF binaries possessed a unique file hash name. To extract the version information from these binaries, we devised a straightforward Python script that scours each ELF for occurrences of "BusyBox v" using the command: \texttt{{"strings \$busybox\_file | grep 'BusyBox v'"}}. Additionally, we employed a regex-based version pattern-matching technique for extracting the version information. The 80 binaries identified were used as the dataset for our fuzzing experiments and crash reuse analysis. 

\subsection{Leveraging LLMs for Initial Seed Generation}
\label{sub-teq1}

We focus our attention on fuzzing the \texttt{awk} applet within the identified BusyBox images as our primary target due to its potential for exploitation. This applet is lightweight interpreter for the \texttt{awk} scripting language, which is often employed in embedded systems to facilitate text processing tasks such as text filtering, pattern matching, and data manipulation. \texttt{awk} scripts may process external input without proper validation, making them susceptible to script injection attacks if the input is not sanitized effectively. Vulnerabilities within these \texttt{awk} scripts can lead to unintended data manipulation or disclosure, posing significant security risks. Because \texttt{awk} takes as its input an \texttt{awk} script, which must conform to a particular language grammar, it is a natural choice of target for LLM-based seed generation, as the LLM can be utilized to easily generate conformant \texttt{awk} scripts. While the brunt of our focus was on \texttt{awk}, LLM-based seed generation is by no means limited to this target; we conduct additional experiments on other target software components to demonstrate this in Section \ref{sec:applicability}. 

\subsubsection{Execution Environment}

Binaries compiled for x86 were evaluated on Ubuntu x86\_64, and ARM-based binaries were evaluated in the QEMU emulator. Notably, we encountered challenges in addressing ARM-specific dependencies, which were effectively resolved by accessing the required dependency files from the company's database. The company's platform had previously extracted the complete filesystem of the target binary, which included the requisite dependency files. This resource proved invaluable in overcoming the challenges associated with ARM target dependencies.

Fuzzer parameters are meant to be user-defined and include the initial input corpus, AFL environment variables to be set, and the fuzzer termination criteria. These criteria include statistical metrics such as runtime, the number of crashes, the number of cycles, and other relevant factors.  Once the fuzzer was configured and initiated, it continuously monitored the specified criteria for termination, halting the fuzzing process upon reaching the defined conditions or in the event of catastrophic errors (conditions that cause AFL++ itself to crash). The outcomes of the fuzzing process were stored separately and categorized by their respective targets, to facilitate subsequent analysis. Failed targets were flagged for further examination and diagnosis. Additionally, JSON dumps of the statistics files corresponding to each target were collected in a shared directory, enabling easy comparison and analysis. Figure \ref{fig:automation-framework} provides an overview of the entire workflow within this framework, illustrating the sequence of steps from dependency management to fuzzing and result collection. The automation script is available at link - \url{https://github.com/asmitaj08/FuzzingBusyBox_LLM}

\begin{figure}[t]
	\centering
	\includegraphics[width=0.8\columnwidth]{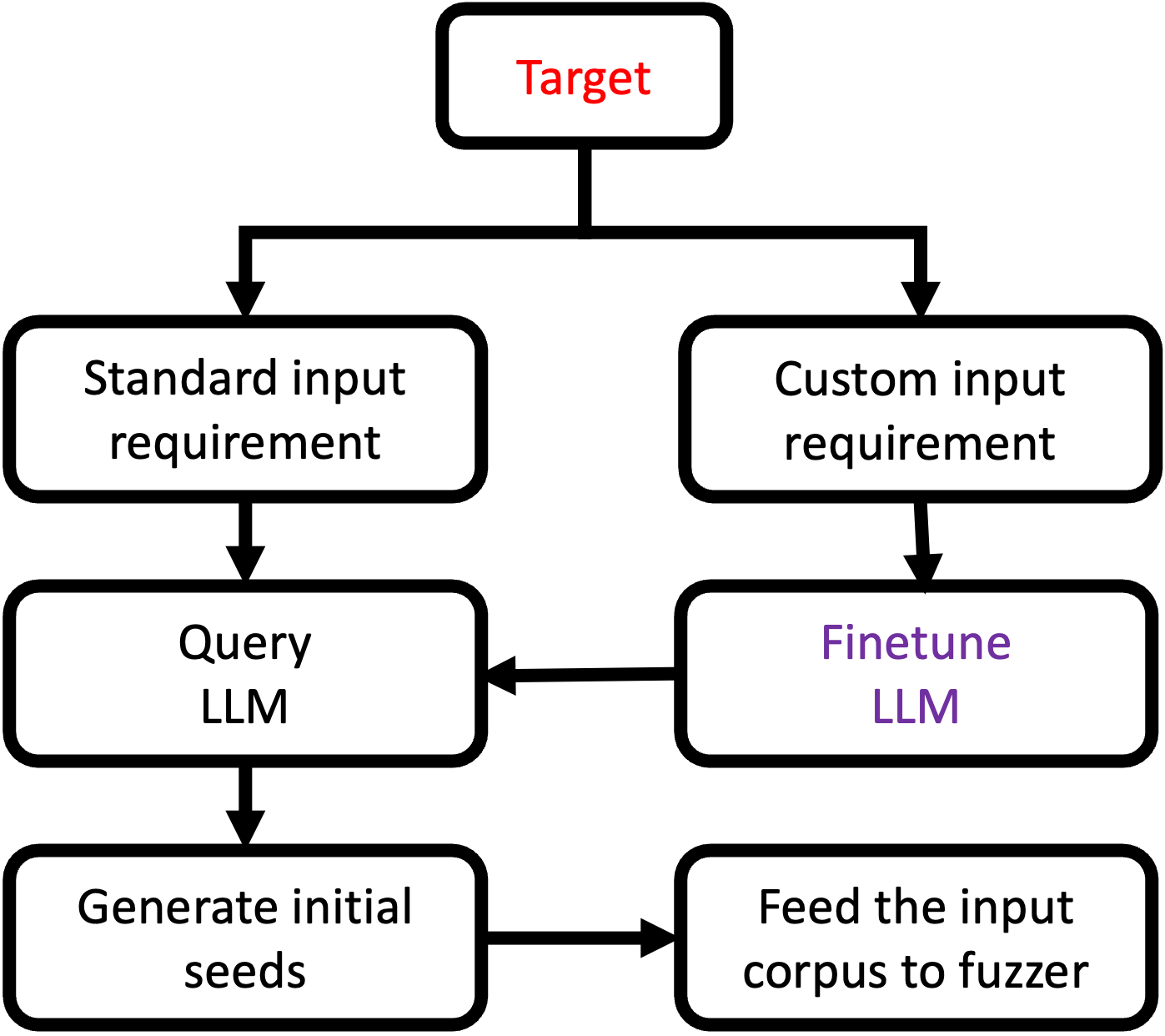}
	\caption{Initial seed generation using LLM}
	\label{fig:openAI-assist}
\end{figure}

\subsubsection{LLM-based Seed Generation Pipeline}

Figure \ref{fig:openAI-assist} visually represents our approach to generating initial seeds for fuzzing through LLM. 

There are 2 scenarios under which initial seeds need to be generated: when the target input format is well-defined and/or standardized, and when the input format is loosely-defined or unknown. When the input format is well-defined, as would is the case for well-known programs like some BusyBox applets, we reason that LLM should not require additional training, as it already possesses knowledge of the expected input format through its initial training on the internet. This can also be determined empirically. However, when the input format of the target is ill-defined or unknown, as would be the case for custom communication protocols, LLM would require fine-tuning. In this scenario, LLM needs to be initially trained with known samples to develop an understanding of the expected input format. In the case of the BusyBox \texttt{awk} applet, we reason that GPT-4 should already be aware of the input format given the applet's popularity. Hence, we did not apply fine-tuning. 

For seed generation, we utilized OpenAI’s GPT model "gpt-4-0613", a chat completion model with a temperature setting of 0.7 (chosen empirically), which we access via queries to the web API. We provided the following prompt to guide the seed generation process for \texttt{awk}:

\textit{"role": "system", "content": "You are initial seed generator for a fuzzer that has to fuzz BusyBox awk applet. In response only provide the list of awk scripts"\\
"role": "user", "content": f"Generate initial seed to fuzz BusyBox awk applet"}

The model responded with a list of commands relevant to the BusyBox \texttt{awk} applet. These commands were then translated into individual \texttt{.awk} scripts, which were subsequently integrated into the input corpus. This input corpus served as the set of initial seeds for the fuzzing process. We used \texttt{afl-cmin} to minimize the input corpus before sending it to the fuzzer, which filters the LLM-generated input corpus to include only the seeds that are useful for fuzzing.

\subsection{Crash Reuse}
\label{sub-teq2}

\begin{figure}[t]
	\centering
	\includegraphics[width=\columnwidth]{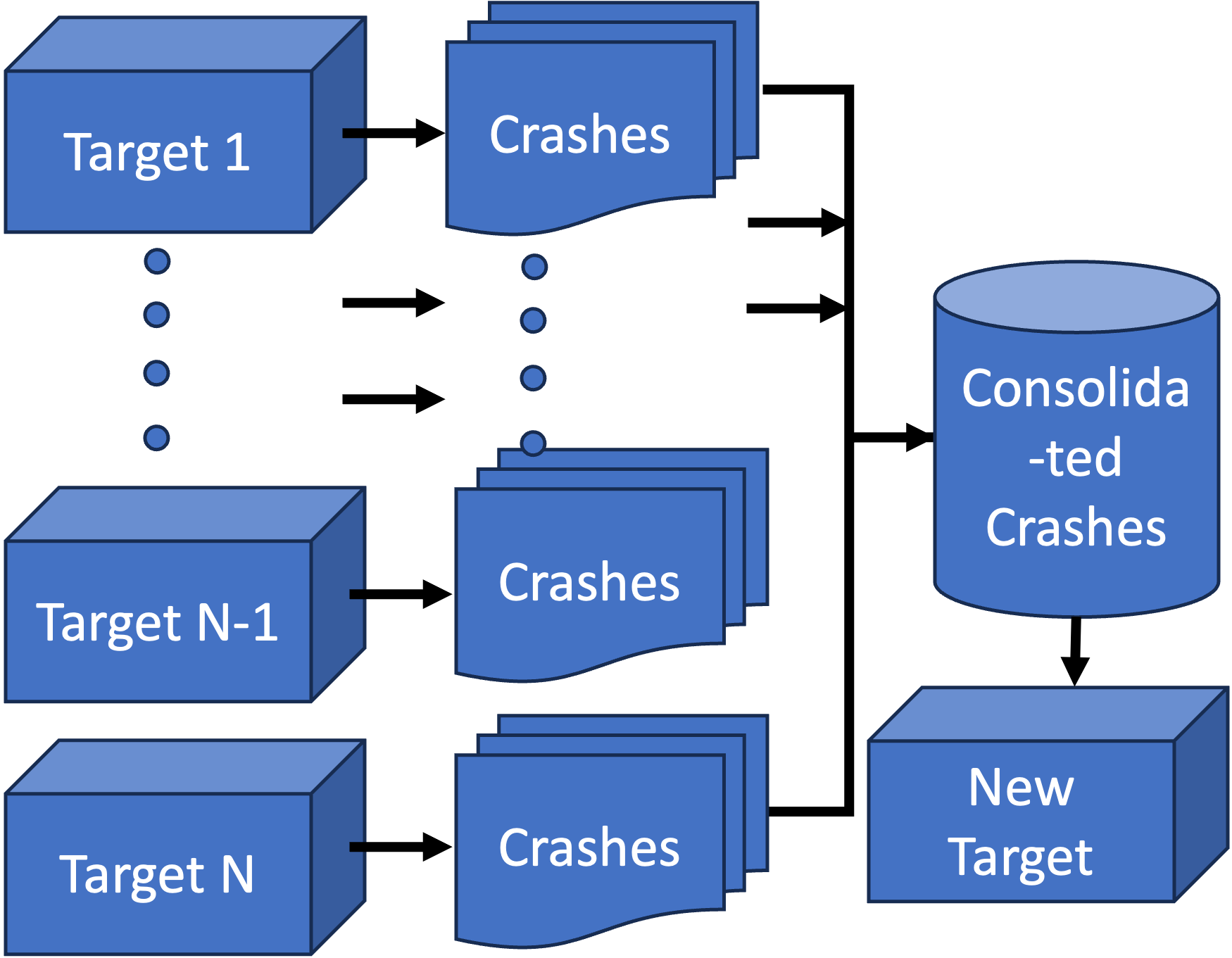}
	\caption{Testing new targets with the existing crash database.}
	\label{fig:crash-based-test}
\end{figure}

Having completed our fuzz testing runs on individual BusyBox targets, we turn our attention to triaging crashes and investigating the potential utility of crash reuse. To recap, we have hypothesized that we can leverage known crashing inputs for a given target to quickly determine if variants of that target contain a similar vulnerability or bug.
As a sanity check, we first evaluate our hypothesis on our set of fuzzed targets by cross-validating the crashing inputs of each target on each other target.

Crash Reuse provides several advantages in software testing: \\
\textit{1. Efficiency:} Initially testing the new target against the consolidated crash database offers the potential for significant time and resource savings. By capitalizing on the crashes identified during previous fuzz testing on similar targets, we can leveraging the resources previously expended in fuzzing and accelerate the fuzzer's coverage exploration by including it in future seeds. Hence, we can potentially identify previously discovered crashes in the new variant without extensive fuzz testing. \\
\textit{2. Black-Box Testing:} This technique is highly beneficial when conducting black-box testing on new variants of a previously tested target. It is particularly advantageous in scenarios where the target utilizes accessible or open-source software components, even if further details are unavailable. By fuzzing open-source variants, we can gather crashing inputs to use as high-quality seeds that are likely to identify duplicate vulnerabilities. This is preferable to engaging in resource-intensive binary-only black-box fuzzing, which can be extremely difficult depending on the complexity of the system under test.

Figure \ref{fig:crash-based-test} provides a visual representation of our approach to crash reuse. We actively curate a database of crashes obtained from previously fuzzed software components. Then, when we encounter variants of these software components in the future, we leverage the collected crashes to identify potential issues in the new variant under test without fuzzing. This provides us with a rapid initial assessment of the new target, which can later undergo more thorough fuzzing for in-depth inspection. As detailed in Section \ref{sec:applicability}, this technique is applicable to any target whose variant has undergone previous fuzzing, and for which we possess a corresponding collection of crashes.

\subsection{Evaluating a New Target}
\label{sub-crash-triage}

After encountering a substantial number of crashes within the collected versions of BusyBox during our research, we applied our techniques to the latest version (1.36.1); this served as an evaluation of our techniques for a new target, as we had collected no samples that contained the most recent BusyBox version. We built our target by compiling the BusyBox from its source code for x86\_64, following the prescribed instructions. We opted not to inject AFL++ instrumentation into the binary.

Our approach to testing the latest BusyBox version was executed in two distinct stages to leverage both techniques effectively:\\
\textit{Stage 1: Crash Reuse - }
In this stage, we applied Technique 2, that is, crash reuse. This technique involved testing the latest version against all the previously obtained crashes from our research without subjecting it to additional fuzzing. The goal was to determine whether some existing crashes can crash the latest version of BusyBox, i.e., increasing the scope of finding vulnerabilities without hours of fuzzing.\\
\textit{Stage 2: Fuzz Testing} - In the second stage, we fuzz tested the target in AFL-QEMU mode using initial seeds generated by GPT-4 with the aim of uncovering additional potential crashes and vulnerabilities. We performed 10 hours of fuzzing on the latest BusyBox on x86\_64 host machine running Ubuntu 22.04. 

\begin{figure}[t]
	\centering
	\includegraphics[width=\columnwidth]{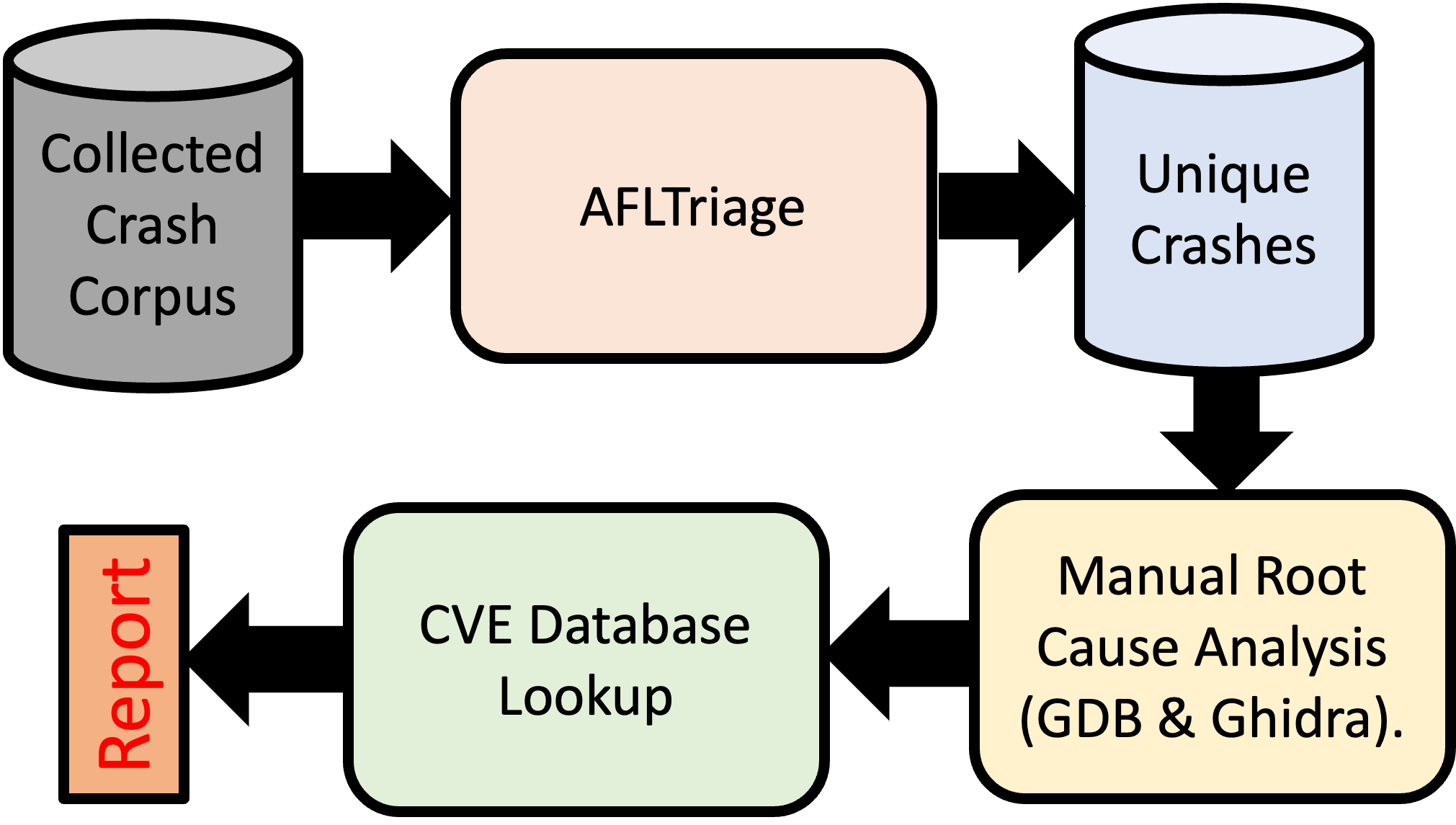}
	\caption{Crash triaging process.}
	\label{fig:crash-triage}
\end{figure}

After our testing stages, we analyzed the collected crashes to identify unique ones, determine the underlying causes, and ascertain whether similar issues had been previously documented. To our knowledge at the time of our research, there was no fully automated tool capable of reliably and comprehensively analyzing fuzzer-induced crashes, making tool-assisted manual analysis the most reliable method of investigation. We performed our analysis using Ghidra \cite{GhidraGithub}, GDB (GNU Debugger) \cite{GDB}, and AFL-Triage \cite{AFLTriage}. We started our analysis with AFL-Triage, which utilizes GDB to triage crashing input files. It categorizes crashes based on their type and reports the debugger's output, making it easier to identify the cause of the crash. It also supports crash deduplication, thus assisting in identifying unique crashes.

Once we obtained the list of unique crashes, we analyzed the input that caused each crash and attempted to minimize it. Moreover, we conducted manual reverse engineering using Ghidra and GDB to identify the root cause. It was followed by searching the CVE (Common Vulnerabilities and Exposures) database to identify similar bugs and conducting further analysis to determine if the bug was because of BusyBox or other libraries on which it was dependent. Figure \ref{fig:crash-triage} shows the overview of the triaging process.

\section{Results}
\label{sec:results}
This section provides an overview of the results obtained through the exploration of the aforementioned techniques.

\subsection{BusyBox Versions in Real-World Embedded Devices}

\begin{table*}
	\centering
    \caption{BusyBox versions in real-world embedded devices}
	\label{tab:busybox-version}
    \vspace{1ex}
	\begin{minipage}{\textwidth}
		\centering
        \resizebox{\textwidth}{!}{%
    		\normalsize
\begin{tabular}{|c|c|c|c|c|c|c|c|c|}
\hline
\begin{tabular}[c]{@{}c@{}}BusyBox \\ Version\end{tabular} & \begin{tabular}[c]{@{}c@{}}No. of \\ Occurrence\end{tabular} & \begin{tabular}[c]{@{}c@{}}Product\\ Types\end{tabular} & \begin{tabular}[c]{@{}c@{}}BusyBox \\ Version\end{tabular} & \begin{tabular}[c]{@{}c@{}}No. of \\ Occurrence\end{tabular} & \begin{tabular}[c]{@{}c@{}}Product\\ Types\end{tabular} & \begin{tabular}[c]{@{}c@{}}BusyBox \\ Version\end{tabular} & \begin{tabular}[c]{@{}c@{}}No. of \\ Occurrence\end{tabular} & \begin{tabular}[c]{@{}c@{}}Product\\ Types\end{tabular} \\ \hline
v1.7.2 & 2 & wireless access point, & v1.19.4 & 4 & \begin{tabular}[c]{@{}c@{}}network management tool, \\ wireless access point, \\ network hardware, \\ security camera\end{tabular} & v1.27.2 & 1 & power distribution unit \\ \hline
v1.10.2 & 1 & telecom device & v1.20.2 & 2 & security camera, & v1.28.3 & 1 & network management tool \\ \hline
v1.11.1 & 1 & building automation & v1.21.1 & 5 & \begin{tabular}[c]{@{}c@{}}drone, ip phone, \\ medical device, bmc\end{tabular} & v1.28.4 & 1 & operating system \\ \hline
v1.13.2 & 1 & building automation & v1.22.1 & 7 & \begin{tabular}[c]{@{}c@{}}wireless aceess point, \\ network switch, \\ operating system\end{tabular} & v1.29.3 & 1 & operating system \\ \hline
v1.15.2 & 2 & wireless aceess point & v1.23.0 & 3 & wireless access point, & v1.30.1 & 6 & \begin{tabular}[c]{@{}c@{}}wireless access point, \\ building automation, \\ power managemnet system\end{tabular} \\ \hline
v1.17.2 & 1 & router & v1.23.1 & 7 & bmc, router, & v1.33.0 & 1 & power management system \\ \hline
v1.17.3 & 1 & telecom device & v1.24.1 & 7 & \begin{tabular}[c]{@{}c@{}}wireless access point, \\ network switch,\end{tabular} & v1.34.0 & 1 & network controller card \\ \hline
v1.17.4 & 1 & printer & v1.25.0 & 3 & \begin{tabular}[c]{@{}c@{}}drone, \\ network attached storage\end{tabular} & v1.34.1 & 4 & \begin{tabular}[c]{@{}c@{}}building automation, \\ wireless access point\end{tabular} \\ \hline
v1.18.2 & 1 & wireless aceess point & v1.25.1 & 3 & wireless access point, & v1.35.0 & 1 & router \\ \hline
v1.19.2 & 1 & wireless aceess point & v1.26.2 & 6 & \begin{tabular}[c]{@{}c@{}}power management system, \\ building automation,\end{tabular} & v1.36.0 & 1 & router \\ \hline
\end{tabular}
        }
	   \end{minipage}
\end{table*}

As discussed in Subsection \ref{sub-version}, our first objective is to shed light on the older versions of BusyBox that are still in use within real-world embedded products. Table \ref{tab:busybox-version} presents an overview of the versions of BusyBox that we discovered in approximately 80 embedded products spanning various categories, including wireless access points, telecom devices, building automation systems, routers, printers, power distribution units, and others. Our findings revealed that many of these devices continue using significantly older BusyBox versions. Notably, the latest version of BusyBox, as of the time of writing this paper, is v1.36.1. However, Table \ref{tab:busybox-version} illustrates that many older versions are still in use. This discovery is concerning, especially given the well-documented vulnerabilities associated with these older versions. 

It is important to emphasize that our investigation focused on a limited set of firmware samples. Considering the vast array of embedded devices that populate the modern landscape, this situation raises significant concerns that demand attention and remediation.
In summary, the outcomes underscore the pressing need for increased awareness and action concerning the usage of outdated and vulnerable versions of BusyBox in real-world embedded devices. A similar situation could arise with other software components.

\subsection{Leveraging LLMs for Initial Seed Generation}

\begin{table}[t]
	\centering
    \caption{Comparison of number of crashes with and without LLM}
	\label{tab:w-wo-LLM-comparison}
    \vspace{1ex}
	\begin{minipage}{\columnwidth}
		\centering
        \resizebox{\columnwidth}{!}{%
    		\normalsize
\begin{tabular}{|l|c|c|c|l|c|c|c|}
\hline
\multicolumn{1}{|c|}{\begin{tabular}[c]{@{}c@{}}Target\\ Version\\ (ARM)\end{tabular}} &
  \begin{tabular}[c]{@{}c@{}}Product \\ Type\end{tabular} &
  \textbf{\begin{tabular}[c]{@{}c@{}}No. of \\ crashes\\ w/o LLM\end{tabular}} &
  \textbf{\begin{tabular}[c]{@{}c@{}}No. of \\ crashes\\ with LLM\end{tabular}} &
  \multicolumn{1}{c|}{\begin{tabular}[c]{@{}c@{}}Target\\ Version\\ (x86\_64)\end{tabular}} &
  \begin{tabular}[c]{@{}c@{}}Product \\ Type\end{tabular} &
  \textbf{\begin{tabular}[c]{@{}c@{}}No. of \\ crashes\\ w/o LLM\end{tabular}} &
  \textbf{\begin{tabular}[c]{@{}c@{}}No. of \\ crashes\\ with LLM\end{tabular}} \\ \hline
v1.34.1 &
  \begin{tabular}[c]{@{}c@{}}embedded \\ wireless \\ controller\end{tabular} &
  3 &
  188 &
  v1.23.1 &
  \begin{tabular}[c]{@{}c@{}}network \\ controller\end{tabular} &
  64 &
  140 \\ \hline
v1.29.3 &
  \begin{tabular}[c]{@{}c@{}}medical \\ device\end{tabular} &
  54 &
  82 &
  v1.22.1 &
  \begin{tabular}[c]{@{}c@{}}network \\ management \\ tool\end{tabular} &
  43 &
  165 \\ \hline
v1.34.1 &
  \begin{tabular}[c]{@{}c@{}}embedded \\ PLC\end{tabular} &
  3 &
  177 &
  v1.30.1 &
  \begin{tabular}[c]{@{}c@{}}network \\ management \\ tool\end{tabular} &
  147 &
  229 \\ \hline
v1.15.3 &
  \begin{tabular}[c]{@{}c@{}}building \\ automation\end{tabular} &
  220 &
  404 &
  v1.27.2 &
  \begin{tabular}[c]{@{}c@{}}operating \\ system\end{tabular} &
  0 &
  114 \\ \hline
v1.23.2 &
  camera &
  50 &
  165 &
  v1.23.1 &
  \begin{tabular}[c]{@{}c@{}}storage \\ array \\ controller\end{tabular} &
  38 &
  178 \\ \hline
v1.18.4 &
  plc &
  137 &
  224 &
  v1.21.1 &
  firewall &
  44 &
  99 \\ \hline
v1.30.1 &
  \begin{tabular}[c]{@{}c@{}}operating \\ system\end{tabular} &
  49 &
  106 &
  v1.19.4 &
  \begin{tabular}[c]{@{}c@{}}network \\ switch\end{tabular} &
  49 &
  172 \\ \hline
v1.26.2 &
  \begin{tabular}[c]{@{}c@{}}security \\ camera\end{tabular} &
  55 &
  70 &
  v1.15.1 &
  \begin{tabular}[c]{@{}c@{}}operating \\ system\end{tabular} &
  166 &
  357 \\ \hline
v1.32.0 &
  \begin{tabular}[c]{@{}c@{}}power \\ control \\ system\end{tabular} &
  0 &
  70 &
  v1.23.1 &
  \begin{tabular}[c]{@{}c@{}}network \\ controller\end{tabular} &
  41 &
  98 \\ \hline
v1.27.2 &
  drone &
  34 &
  147 &
  v1.35.0 &
  \begin{tabular}[c]{@{}c@{}}network \\ management \\ tool\end{tabular} &
  2 &
  193 \\ \hline
\end{tabular}
        }
	   \end{minipage}
\end{table}
\begin{figure*}[t]
	\centering
	\includegraphics[width=\textwidth]{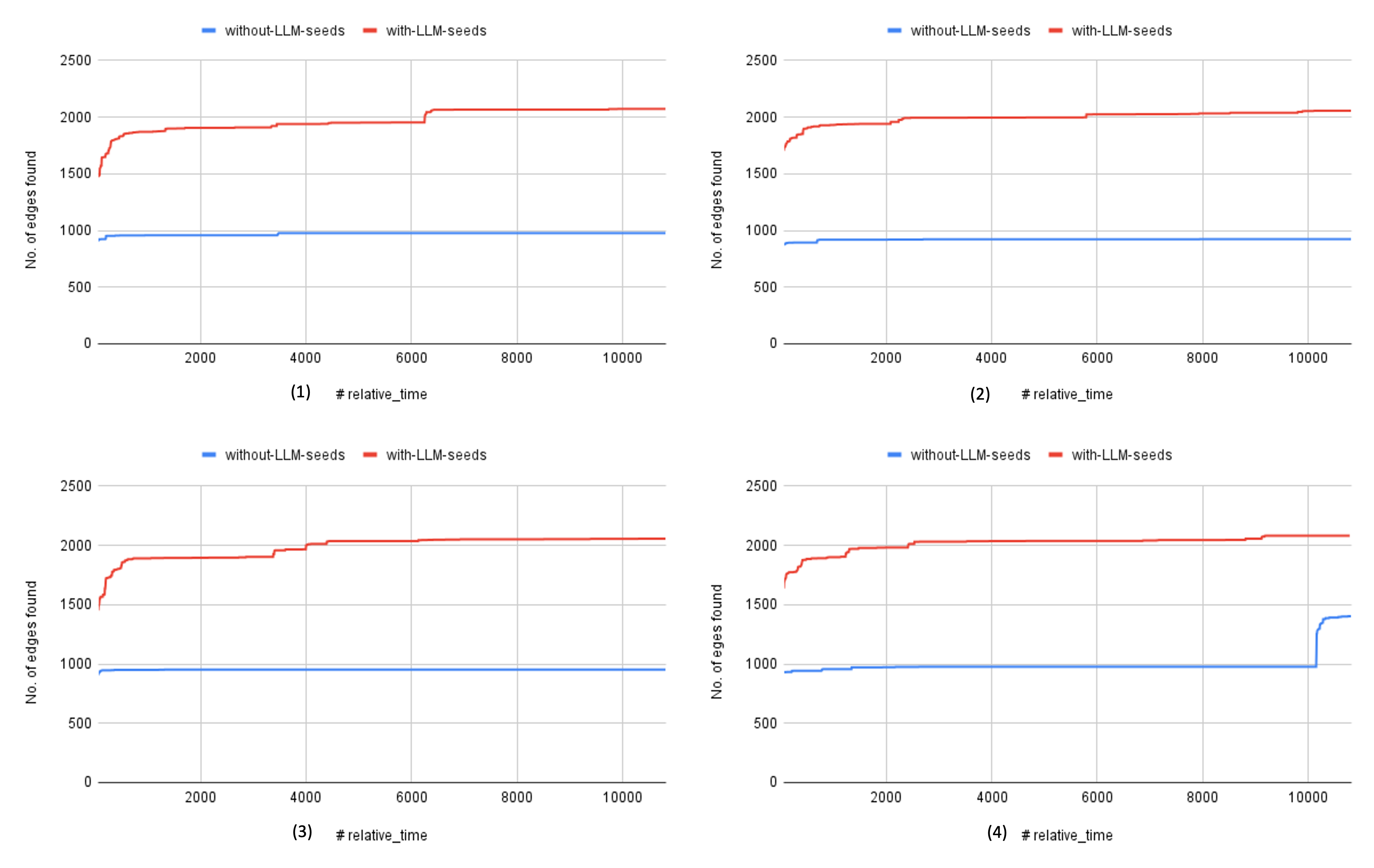}
	\caption{Comparison of number of edges covered with and without using LLM for initial seed generation. Targets are BusyBox in (1) Network controller, (2) Network switch, (3) Storage array controller, (4) Firewall}
	\label{fig:graph_no_of_edges_LLM_teq1}
\end{figure*}

\begin{figure}
	\centering
	\includegraphics[width=\columnwidth]{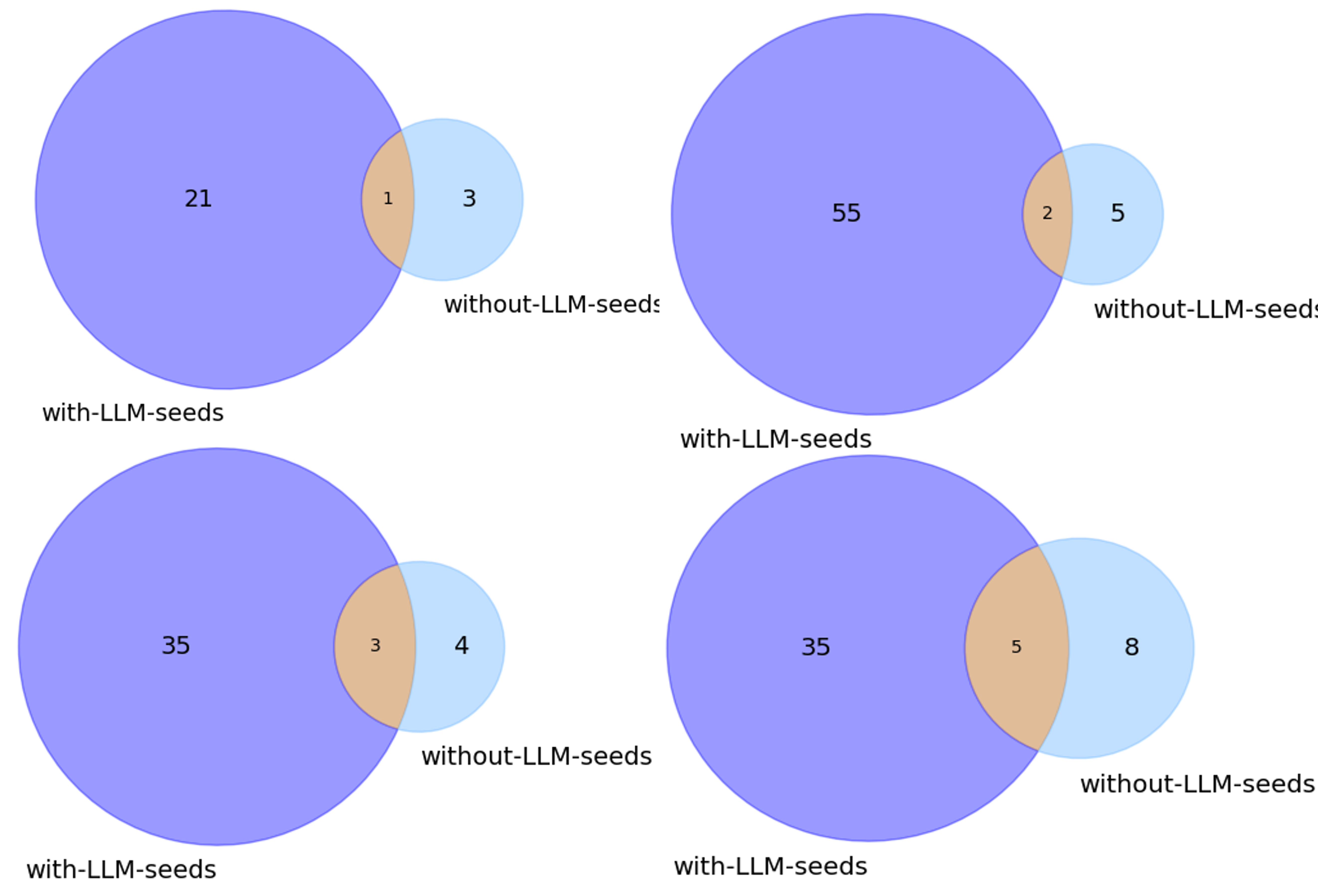}
    	\caption{Comparison of number of unique crashes found with and without using LLM for initial seed generation. Targets are BusyBox in (1) Network controller, (2) Network switch, (3) Storage array controller, (4) Firewall}
	\label{fig:venn-diag-unique-crashes}
\end{figure}

Initially, we executed the AFL++ fuzzer for 3 hours on each BusyBox AWK applet target, utilizing the default AFL++ settings in QEMU mode. This means that no AFL++ instrumentation was applied during compilation. During this phase, we recorded the number of crashes that occurred in each target as well as the number of edges identified throughout the fuzzing process. Then, as detailed in Subsection \ref{sub-teq1}, we employed GPT-4 to generate the initial seeds. In this phase, we repeated the 3-hour fuzzing process for each target. We repeated the same metrics analysis, counting the number of crashes and edges discovered for each target.

The results from some of the ARM\_32 and x86\_64-based BusyBox targets are presented in Table \ref{tab:w-wo-LLM-comparison}. It illustrates that significantly more crashes were identified during fuzzing when the initial seeds were generated by LLM compared to the ones with random seeds. Moreover, the same pattern was observed in the case of the number of edges found during the fuzzing of each of the targets. To visually represent this, Figure \ref{fig:graph_no_of_edges_LLM_teq1} provides a graph displaying data from four targets, contrasting the number of edges discovered in the two scenarios, i.e., initial seeds generated by LLM versus random seeds. \textit{Note: In the tables and figures, the term "without-LLM" signifies scenarios where random initial seeds were used, while "with-LLM" denotes cases where initial seeds were generated using LLM, and ‘relative\_time` is the running time of the fuzzer in seconds}. 

Furthermore, Figure \ref{fig:venn-diag-unique-crashes} presents a Venn diagram depicting the number of unique crashes found in each case and the number of crashes common in both cases. This graphic emphasizes the importance of discovering a more significant number of crashes. When there are more crashes to work with, there are more opportunities to discover different failing execution paths, thereby increasing the likelihood of uncovering vulnerabilities. Figure \ref{fig:venn-diag-unique-crashes} underscores this by revealing that more unique crashes were identified when utilizing LLM-generated initial seeds. However, comprehensive triaging was not carried out for the older versions of BusyBox, and such triaging was primarily focused on the latest version. Hence, in the discussed cases here, we conducted fuzzing, collected crashes, used AFL-Triage to categorize them, and recorded the unique crashes, notably more abundant in the LLM-generated initial seed scenarios.

\begin{table}[t]
	\centering
    \caption{Work leveraging LLM for fuzzing}
	\label{tab:existing-LLM-work}
    \vspace{1ex}
	\begin{minipage}{\columnwidth}
		\centering
        \resizebox{\columnwidth}{!}{%
    		\normalsize
\begin{tabular}{|c|c|c|}
\hline
\textbf{Work} &
  \textbf{Target} &
  \textbf{LLM use case} \\ \hline
\begin{tabular}[c]{@{}c@{}}ChatAFL\\ \cite{Meng2024-LargeLanguageModel}\end{tabular} &
  Protocols &
  \begin{tabular}[c]{@{}c@{}}Extract a machine-readable \\ grammar for a protocol,\\ generate diverse messages\\ for initial seeds.\end{tabular} \\ \hline
\begin{tabular}[c]{@{}c@{}}ChatFuzz\\ \cite{Hu2023-AugmentingGreyboxFuzzing}\end{tabular} &
  \begin{tabular}[c]{@{}c@{}}Format conforming\\  targets\end{tabular} &
  \begin{tabular}[c]{@{}c@{}}Used at the mutationg \\ stage to generated \\ format conforming\\ mutated inputs\end{tabular} \\ \hline
\begin{tabular}[c]{@{}c@{}}Fuz4All\\ \cite{Xia2024-Fuzz4AllUniversalFuzzing}\end{tabular} &
  \begin{tabular}[c]{@{}c@{}}Targets that need\\  different programming \\ languages as input\end{tabular} &
  \begin{tabular}[c]{@{}c@{}}Generate code snippets\\ for different programming \\ languages\end{tabular} \\ \hline
\begin{tabular}[c]{@{}c@{}}WhiteFox\\ \cite{Yang2023-WhiteboxCompilerFuzzing}\end{tabular} &
  Compiler &
  \begin{tabular}[c]{@{}c@{}}Optimization source code\\ analyzer, test input\\  generation\end{tabular} \\ \hline
\begin{tabular}[c]{@{}c@{}}Proposed\\ Work\end{tabular} &
  \begin{tabular}[c]{@{}c@{}}Embedded applications \\ like BusyBox\end{tabular} &
  \begin{tabular}[c]{@{}c@{}}Generate diverse and target- \\ specific initial seeds.\end{tabular} \\ \hline
\end{tabular}
        }
	   \end{minipage}
\end{table}

\subsection{Crash Reuse}

\begin{figure}
	\centering
	\includegraphics[width=\columnwidth]{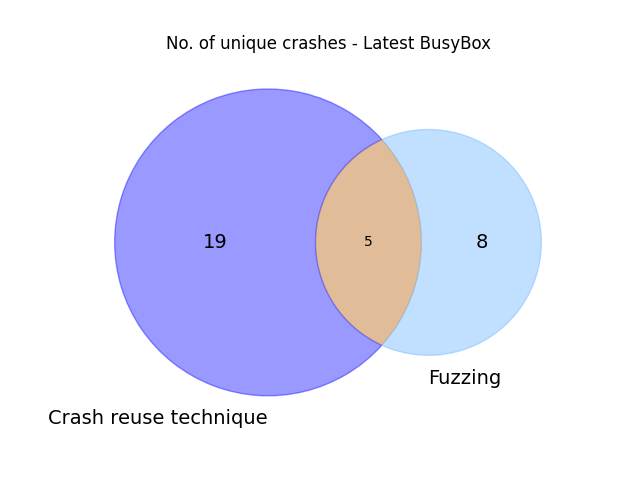}
	\caption{Comparison of number of unique crashes found using crash reuse technique vs fuzzing}
	\label{fig:crash-reuse-result}
\end{figure}

After we had amassed a substantial number of crashes from fuzzed BusyBox targets, our total collection amounted to 4540 crashes that likely map to a much smaller place in the binary where the crash happens. Subsequently, as outlined in Subsection \ref{sub-teq2}, we subjected the latest BusyBox version (v1.36.1) to testing against all these pre-existing crashes. This endeavor discovered 97 crashes in the latest BusyBox, of which 19 were unique. Later, we conducted traditional fuzzing on the latest BusyBox using AFL++ QEMU mode, with initial seeds generated by LLM over 10 hours. This approach yielded 20 crashes, of which eight were unique. Remarkably, five of these eight unique crashes were also identified using the crash reuse technique. Figure \ref{fig:crash-reuse-result} presents a graphical comparison of the number of unique crashes discovered using the crash reuse technique versus traditional fuzzing, as well as the common crashes between the two methods. 

These results underscore the potential utility of crash reuse in software testing. As discussed in Section \ref{sub-teq2}, it can reduce substantial time and resource demands, and is a valuable tool for blackbox fuzzing when a comprehensive crash database is available.

Additionally, it is essential to note that not all crashes indicate software bugs. Crashes can occur for various reasons, including invalid inputs, false positives, unreachable code, execution environment factors, platform-specific issues, and other non-bug-related causes. Reaching conclusive determinations often involves meticulous manual triaging, which can be time-consuming and intricate. As such, we limited our scope to identifying crashes, with triaging performed only on a subset of crashes found in the latest BusyBox version. However, as previously discussed, the quantity of crashes is a vital metric in fuzzing. A higher number of unique crashes equates to a more extensive array of test scenarios to explore during testing. Consequently, this increases the likelihood of identifying potential software vulnerabilities or bugs.

\subsection{Crash Analysis: Latest BusyBox (v1.36.1)}

Following the collection of crashes for the latest BusyBox (v1.36.1) target using LLM-based seed generation and Crash Reuse, we proceeded with manual crash triaging. Due to this process's intensive time and resource commitments, we limited our triaging efforts to 15 unique crashes that resulted in segmentation faults. Crashes obtained from fuzzing can often appear disordered and incomprehensible due to data randomization through various mutation strategies during the fuzzing process. We attempted to minimize the crash size to facilitate the triaging process, making it more comprehensible.

Additionally, it is important to note that BusyBox relies on GLIBC (GNU C), which provides standard C library functions and system calls for Unix-like operating systems. During our triaging endeavor, we uncovered specific input patterns that triggered crashes in various functions within the GLIBC library. These patterns were identified by tracing the crash causes using GDB. Out of the 15 crashes we triaged, we found crashes in GLIBC functions, including \texttt{free, malloc, write, strlen, strdup, regex, and strftime}. Notably, the crashes in \texttt{regex, and strftime} closely resembled the known bugs documented as \texttt{CVE-2010-4051} and \texttt{CVE-2015-8776}, respectively. Although these CVEs were identified quite some time ago, we encountered them in GLIBC versions 2.35 and 2.38 on a host running Ubuntu 22.04, and these issues were also reproducible in Debian distributions. 
It highlights the persistence of these vulnerabilities across multiple software component versions, necessitating renewed attention and remediation efforts. We subsequently filed bug report for these findings; though the bugs appear to have not been considered possibly because it is same as the bug whose CVE has already been assigned. In addition to the crash input patterns chosen for triaging that triggered crashes in GLIBC, other crash inputs also induced segmentation faults within BusyBox. After manual triaging, it became apparent that many of these crashes were primarily attributed to improper memory access. We could not discover any readily exploitable bugs. Stead, we identified crash patterns that, after more in-depth exploration, could possibly reveal vulnerabilities.

\subsubsection{Crash details} \label{vuln-details}

\begin{figure}
    \centering
    \includegraphics[width=\columnwidth]{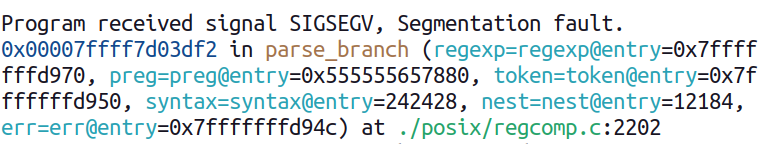}
    \caption{Segmentation fault in regcomp}
    \label{fig:gdb-stack-exhaust-seg}
\end{figure}

The analysis of crashes was done manually using GDB and Ghidra.The issue identified in \texttt{regex} was the denial of service (DoS) caused by memory exhaustion. The pattern, a long repeated character, triggered deep recursion that caused stack exhaustion, leading to a segmentation fault. The crash pattern was \texttt{/1((((.......12208 times/1}. We passed this crash pattern to Busybox (v1.36.1) awk and traced via gdb using \texttt{gdb --args busybox awk -f crash\_pattern\_file}. It caused \textit{segmentation fault} as shown in Figure \ref{fig:gdb-stack-exhaust-seg}.

\begin{figure}
    \centering
    \includegraphics[width=\columnwidth]{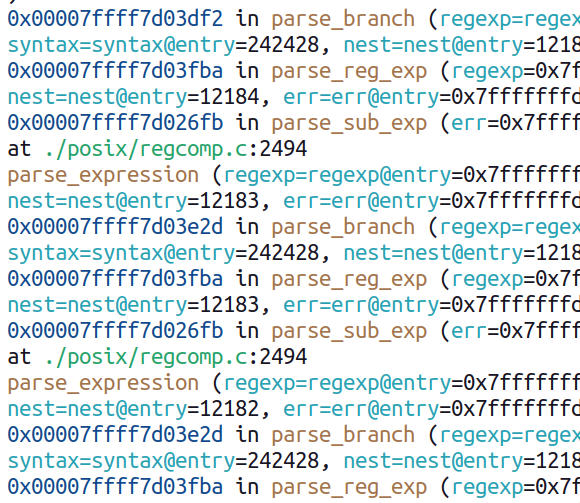}
    \caption{Deep recursion in regcomp }
    \label{fig:gdb-stack-exhaust-bt}
\end{figure}

\begin{figure*}
    \centering
    \includegraphics[width=0.8\textwidth]{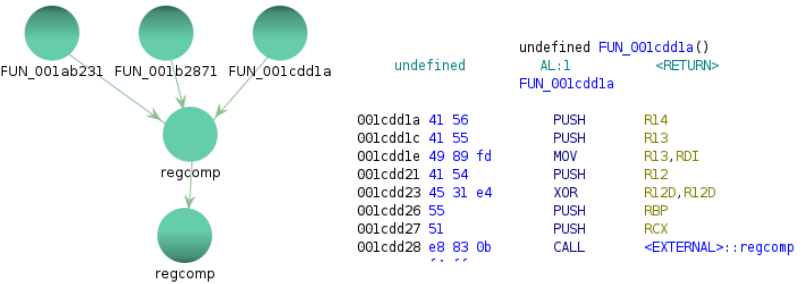}
    \caption{Regcomp called inside Busybox}
    \label{fig:regcomp-callgraph-busybox}
\end{figure*}

Having identified the segmentation fault, we inspected the register values using gdb commands followed by doing backtrace (\texttt{bt}). The outcome of \texttt{bt} showed the issue of deep recursion leading to stack exhaustion as shown in Figure \ref{fig:gdb-stack-exhaust-bt}, which can lead to denial of service (DoS). We further verified it by installing the latest version of GLIBC from the source code and confirming reproducibility. This vulnerability is nearly identical to \texttt{CVE-2010-4051}, which according to Red Hat \cite{RedHatProductSecurity2010-CVE20104051} is due to a failure to consider crash of client application via regcomp. Unfortunately, even on the client side where it is used, the provided pattern is not verified. In this scenario, the pattern being sent to \texttt{regcomp()} does not get verified beforehand, as shown in Figure \ref{fig:regcomp-callgraph-busybox}.

\begin{figure}
    \centering
    \includegraphics[width=\columnwidth]{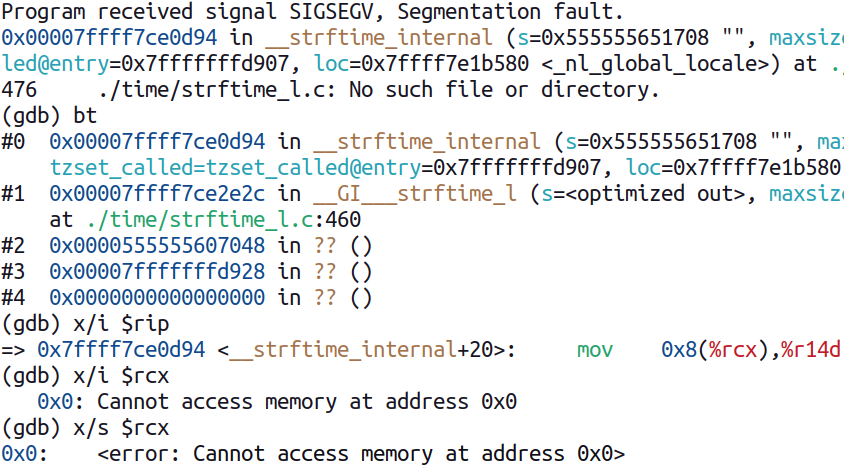}
    \caption{Crash in strftime because of invalid args send via BusyBox awk}
    \label{fig:strftime-bt}
\end{figure}

Using a similar approach to the one discussed above, we analyzed a number of other crashes. One of the other crashes was found in \texttt{strftime} was also DoS because of invalid pointer to \texttt{struct tm}. The crash pattern sent to BusyBox \texttt{awk} applet was \texttt{BEGIN{strftime("", "3333333333333333333")}}, leading to a segmentation fault caused by \texttt{\_\_strftime\_internal()}, as shown in Figure \ref{fig:strftime-bt}. Similar to the case of \texttt{regcomp}, \texttt{strftime} is being called within BusyBox, but the the input parameters being sent to it is not being verified beforehand, leading to DoS. Other crashes that were analyzed had resulted from abnormal patterns in the crash input and therefore could not be conclusively identified as software bugs. 

\section{Applicability of proposed techniques}
\label{sec:applicability}
The initial proof of concept targeted the BusyBox \texttt{awk} applet. This section extends the application of the proposed technique to other BusyBox applets found in various older versions used in real-world embedded products, as detailed in Table \ref{tab:busybox-version}. The fuzzing process, conducted over 48 hours. Apart from the number of crashes, we also examined the number of covered edges (as shown for the \texttt{awk} applet in Figure \ref{fig:graph_no_of_edges_LLM_teq1}) and the total number of executions. This assessment involved testing on the \texttt{dc}, \texttt{man}, and \texttt{ash} applets along with \texttt{awk}.

\begin{figure*}[t]
    \centering
    \includegraphics[width=\textwidth]{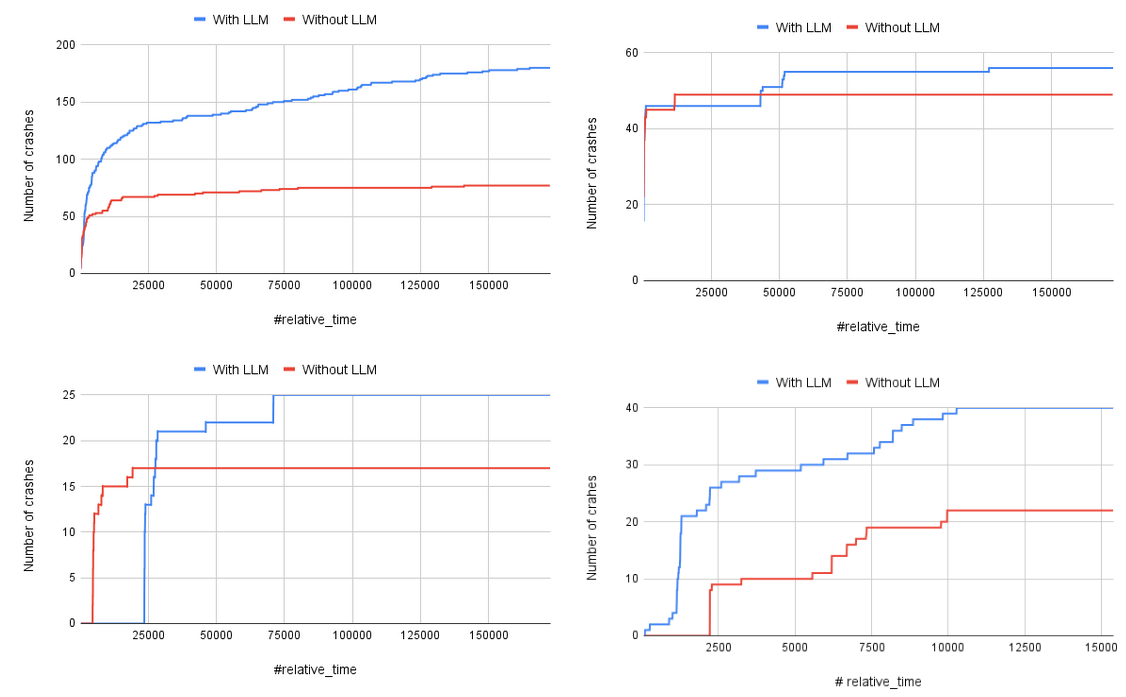}
    \caption{Number of crashes with vs without LLM based initial seeds. The graphs are for applet \texttt{awk}, \texttt{dc}, \texttt{man}, \texttt{ash} clockwise.}
    \label{fig:48h-crashes}
\end{figure*}

\begin{figure*}[t]
    \centering
    \includegraphics[width=\textwidth]{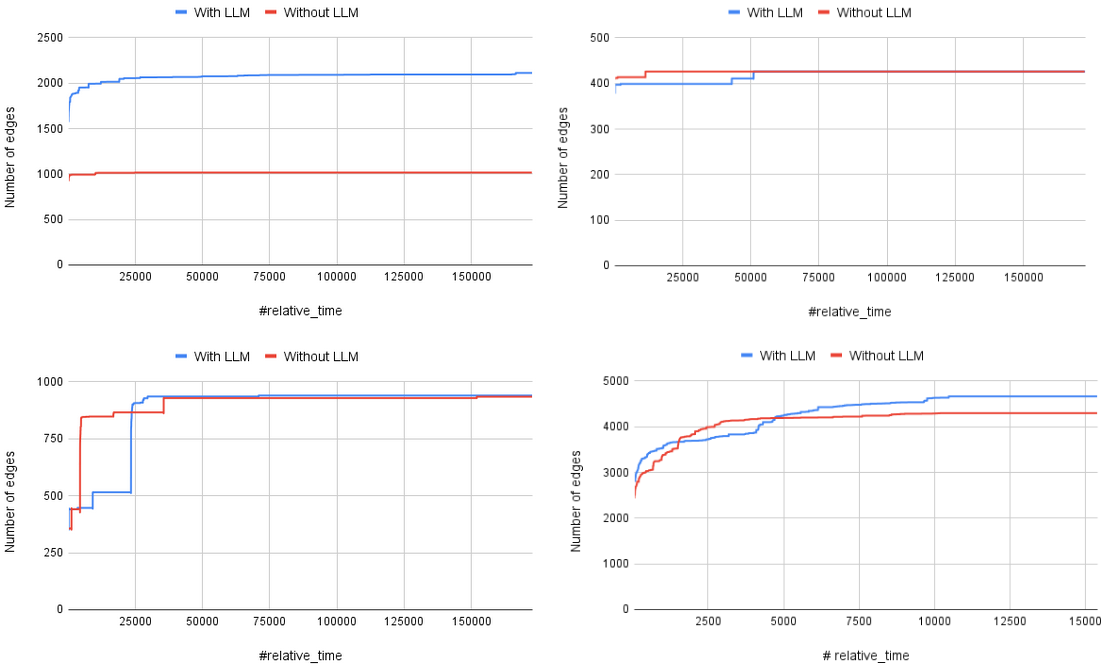}
    \caption{Number of edges covered with vs without LLM based initial seeds. The graphs are for applet \texttt{awk}, \texttt{dc}, \texttt{man}, \texttt{ash} clockwise.}
    \label{fig:48h_edges}
\end{figure*}

\begin{figure*}[t]
    \centering
    \includegraphics[width=\textwidth]{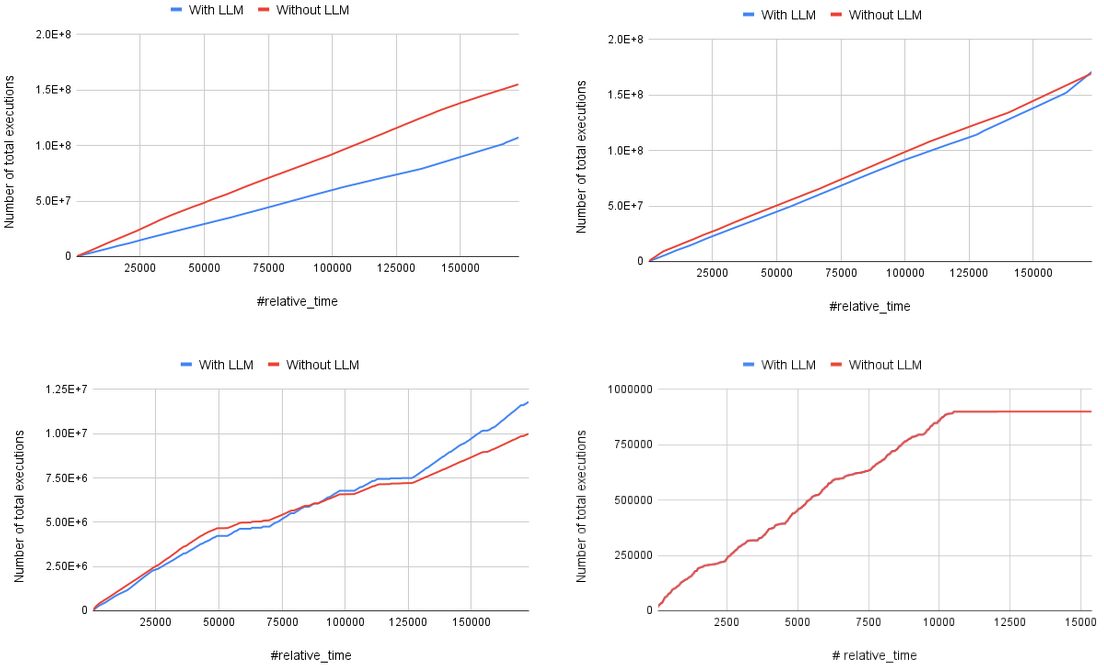}
    \caption{Number of executions with vs without LLM based initial seeds. The graphs are for applet \texttt{awk}, \texttt{dc}, \texttt{man}, \texttt{ash} clockwise.}
    \label{fig:48h_exec}
\end{figure*}

Figure \ref{fig:48h-crashes} illustrates variations in number of crash detected across different targets. Notably, using LLM-generated initial seeds led to a substantial increase in crashes for certain cases like \texttt{awk}, while for \texttt{dc} and \texttt{man}, crashes were decently higher with LLM-generated seeds. For the \texttt{ash} applet, we conducted fuzzing for 5 hours due to utility execution constraints. As a minimized version of \texttt{bash}, \texttt{ash} executes shell commands. However, abnormal behavior during fuzzing, triggered by diverse shell inputs, led to the cessation of the process. Consequently, we opted for a shorter testing duration. This pattern also extends to edge coverage performance as shown in Figure \ref{fig:48h_edges}, with some cases showing higher performance and others exhibiting negligible differences. Furthermore, when considering the number of executions as shown in Figure \ref{fig:48h_exec}, it is evident that LLM-based generated seeds do not introduce significant overhead in most cases. 

The LLM-based technique for generating initial seeds plays a supportive role in the fuzzing process, contributing quality and diverse seeds that enhance fuzzing performance. However, it is crucial to note that this technique alone is not the sole factor influencing the overall outcome. Various associated factors, contingent upon the specific target, must be considered. The effectiveness is contingent upon the target type and the extent to which the initially provided diverse inputs contribute to code coverage. 

The primary function of LLM in this proposed technique is to assist in producing high-quality and diversified initial seeds, thereby potentially enhancing fuzzing performance. The seed generation using LLM requires initial manual intervention to validate if the generated seeds align with the target's requirements. In the case of a new target initially unknown to LLM, model training is essential for the target-specific seed format. However, this represents an initial, one-time effort; once the model learns the required seed format, it expedites the generation of diverse seeds suitable as potential initial seeds for fuzzing. Thus, we can leverage the knowledge base of LLM models, or train these models according to different target requirement. Therefore, this technique is not restricted to the BusyBox and can be adapted for use with different targets.

Similarly, the crash reuse technique proposed can be extended to various software components across different targets. The technique isn't limited to a particular target but applies universally. It can be employed in any scenario where we have previously gathered crash data by fuzzing a target and aim to test the variant of that target by reusing those crashes. For instance, following the collection of crashes from the tested samples of the \texttt{dc} applet, akin to the approach outlined in Section \ref{sub-teq2} for the \texttt{awk} applet, we reused these crashes to test other BusyBox samples. These samples included different versions and architectures. We had a total of 2112 crashes from previously tested samples that were ARM-based BusyBox targets. We reused these crashes to test if they could cause crashes in BusyBox v1.36.1, which is x86-based. Out of 2112 crashes, 853 caused segmentation faults in this new target, with 313 being unique occurrences. Thus, using this technique, we identified the possibility of crashes in a new target even without performing actual fuzzing on it. Due to limited time, we didn't delve deeply into these crash exploration paths. The primary goal is to convey that the crash reuse technique could be beneficial for the initial screening of a new variant of a software component without spending hours on fuzzing. However, this technique may not uncover all the vulnerabilities requiring fuzzing for a thorough analysis.

While crash replay has been established for analyzing crashes in previously fuzzed targets, there is limited existing research on collecting crashes and reusing them to test variant of the software components present in different targets. For instance, different products may incorporate different versions of BusyBox. This distinction becomes crucial in the context of embedded systems, where common third-party software components like BusyBox are frequently shared. However, these components may differ in version numbers, architectures, or compilation optimizations. 

The crash reuse becomes particularly advantageous when we have access to the open-source version of specific software components or libraries. By conducting fuzzing on these components, we can generate potential crashes. Given our access to the source code, the likelihood of identifying these crashes is higher.
Now, let's consider a scenario where the same software component is internally used in a product for which we only have access to the binary, not the source code. Although we are aware that it internally employs the software component, there is a possibility that its version number differs, or it has been customized by the developer, or compiled for a different architecture. In such cases, where we have already collected crashes for the known software component, we can employ this crash-reuse technique as an initial screening to test for the presence of vulnerabilities in the given product.

As illustrated in Figure \ref{fig:pipeline}, our work introduces a novel pipeline employing two techniques. In this pipeline, LLM-generated initial seeds (technique1) assists in enhancing fuzzing and acquiring crashes for the target under test. Subsequently, the obtained crashes are collected and reused (technique2)  to test a new target with similar software components.

\section{Discussion}
\label{sec:discussion}

 In our pursuit to enhance existing software testing methodologies, we emphasize the significance of our proposed techniques, particularly within the context of embedded systems. Firmware in embedded systems often consists of numerous third-party software components with custom implementations and unique input types, making it predominantly a black-box testing scenario. The techniques introduced in this work, namely leveraging LLM for initial seed generation and crash reuse, have exhibited promising outcomes that can significantly aid software testing efforts. While these techniques can be adapted for various targets as discussed in Section \ref{sec:applicability}, we have used them to analyze BusyBox for the proof of concept. 

 To evaluate our results, we established the AFL++ fuzzer's output as our baseline and compared outcomes between AFL++ provided with random initial seeds versus LLM-generated initial seeds. Despite extensive research leveraging Large Language Models (LLM) for fuzzing, as discussed in Section \ref{sec:related} and exemplified by studies like \cite{Huang2024-LargeLanguageModels}, we have not encountered prior work applying LLM to real-world embedded devices running applications like BusyBox. Table \ref{tab:existing-LLM-work} illustrates various studies leveraging LLM for different targets, making direct comparisons unfeasible.

Similarly, regarding the crash reuse technique, we haven't identified any work collecting and reusing crashes to test similar software components on a different target, as depicted in Figure \ref{fig:crash-based-test}. Earlier methodologies, like reFuzz \cite{reFuzz}, focused on reusing crashes and outputs within the same target during different stages of fuzzing. AFL also provides a feature to resume the fuzzer, leveraging previous crashes and queue information to enhance fuzzing outcomes on the same target. Crashes are typically employed for replay during the analysis process. Therefore, for comparison purposes, we set the baseline as identifying the number of crashes directly using fuzzing versus testing the target against the crashes collected from fuzzing the variant of that target previously, as illustrated in Figure \ref{fig:crash-reuse-result} 
 
 Nevertheless, it is essential to acknowledge certain limitations and challenges associated with these approaches. Utilizing LLM for initial seed generation may necessitate a significant initial effort, mainly when dealing with different targets, especially in the complex domain of embedded systems where a wide array of hardware protocols and custom input patterns are encountered. Furthermore, while the crash reuse technique represents a valuable first pass phase, it may not consistently identify all bugs, especially zero-day vulnerabilities. Hence, a traditional fuzzing technique remains a necessary complement for comprehensive testing. The crash reuse method primarily assists in determining whether previously identified crashes are applicable to a new target but does not guarantee the discovery of all potential bugs. Furthermore, there is a prospect for utilizing LLM to generate high-quality crashes by training the model for specific targets using previously identified crashes. We explored the feasibility of fine-tuning LLM by providing sets of crashes corresponding to respective target categories. We aimed to ascertain whether LLM could generate test cases capable of inducing crashes in the new target. However, this undertaking introduced specific challenges, primarily related to data encoding. LLM requires data to be JSON-encoded, and managing data that cannot be UTF-8 encoded proved to be intricate. This challenge is particularly pertinent, as some crash test cases may involve data types that are not UTF-8 encoded.

In summary, while there are particular challenges and limitations, substantial research potential exists for harnessing these techniques to enhance and assist software testing endeavors. These approaches hold promise for improving the efficiency and effectiveness of testing procedures, particularly in the context of embedded systems and firmware analysis.

\section{Future Work}
\label{sec:future}

This work opens up opportunities for further research by extending the application of LLM and crash-reuse techniques to a broader range of targets within embedded firmware. We plan to implement these techniques on various other targets, encompassing application-level targets like web servers, network-related components, and bare-metal embedded targets that interact with hardware and IoT protocols. Training LLM to understand the input structures of protocols such as I2C, SPI, UART, MQTT, Bluetooth, and others could greatly enhance fuzz testing for these devices. Security testing for embedded targets presents numerous challenges \cite{WYCINWYC}, and incorporating these techniques could be invaluable.

Furthermore, many embedded targets lack access to source code, and in some cases, internal details are undisclosed. In such scenarios, the crash reuse technique can be a valuable resource. By testing unknown targets against existing crashes and evaluating whether any input can cause the target to crash, these techniques can significantly improve the state of security testing for firmware. We are dedicated to exploring these techniques further and leveraging their potential to enhance security testing for a wide range of embedded devices in the future.

\section{Conclusion}
\label{sec:conclusion}

In conclusion, our exploration into BusyBox, driven by its extensive presence in Linux-based embedded devices, has yielded valuable insights and introduced techniques to enhance the software testing process.
Our initial investigation revealed the prevalence of older versions of BusyBox in real-world embedded devices, prompting us to delve further into the analysis. This exploration led to the developing of two techniques to bolster software testing efforts. First is, leveraging LLM for initial seed generation. This technique significantly improved the outcome of fuzzing by enhancing the number of identified crashes, offering a more comprehensive and effective testing approach. Second is, crash reuse technique. We leveraged previously obtained crashes in older versions of BusyBox to assess their applicability to the newer version. This approach proved successful when applied to the latest version of BusyBox, saving time and resources in the testing process.

Subsequently, we delved into the analysis, identifying unique crashes using AFLTriage and conducting manual crash triaging using GDB and Ghidra on 15 of these unique crashes. Our triaging efforts uncovered crashes that triggered issues within the GLIBC library, a critical dependency for BusyBox. These issues bore a resemblance to previously documented CVEs, underscoring the persistent nature of these vulnerabilities across different versions of GLIBC. While the exploitability of the crashes found in BusyBox could not be conclusively determined due to time constraints, our exploration of BusyBox has illuminated techniques with significant potential to benefit software testing in various domains. The findings from this research not only shed light on the security landscape of BusyBox but also open the door to further research and advancements in software testing methodologies.Moreover, as discussed in Section \ref{sec:applicability}, the proposed techniques could be applied to targets other than BusyBox.

\section*{Acknowledgment}
\label{sec:ack}
We would like to acknowledge NetRise team for providing us with real-world embedded firmware database and cloud resource to perform a part of these experiments. We would also like to thank NSF CHEST for funding this project \textit{(Project \# 1916741 industry funding)}.





\small
\bibliographystyle{plain}
\bibliography{references.bib}

\end{document}